# Quantum Treatment of Kinetic Alfvén Waves instability in a dusty plasma: Magnetized ions

N. Rubab[1] and G. Jaffer[2]
1. Department of Space science, Institute of Space Technology, Islamabad, Pakistan
2. Space Science Department, University of the Punjab, Lahore, Pakistan

July 20, 2016

## Abstract

The dispersion relation of kinetic Alfvén wave in inertial regime is studied in a three component non-degenerate streaming plasma. A linear dispersion relation using fluid- Vlasov equation for quantum plasma is also derived. The quantum correction $C_Q$ raised due to the insertion of Bohm potential in Vlasov model causes the suppression in the Alfven wave frequency and the growth rates of instability. A number of analytical expressions are derived for various modes of propagation. It is also found that many system parameters, i.e, streaming velocity, dust charge, number density and quantum correction significantly influence the dispersion relation and the growth rate of instability.
Key words: non-degenerate, Bohm potential, dust grains, kinetic Alfven waves

## 1 Introduction

Theoretical and laboratory studies on current-driven electromagnetic instabilities have been a popular subject in plasma research [1, 2]. Electromagnetic and electrostatic plasma instabilities driven by the parallel or cross field currents give rise to narrow banded spontaneous emissions and have received a lot of attention so far. Its has been speculated that low frequency electromagnetic waves, i.e., Alfvén waves are the most important electromagnetic waves which take part in various dynamical processes happening in sun [3]$^-$[5]. Recently, kinetic Alfvén waves (KAWs) have been modified via quantum effects associated with electrons, for example, Hussain et. al., made an attempt to study spin and nonrelativistic quantum effects on KAWs within the frame work of kinetic theory of Alfvén waves, [6] and their analysis suggested that the spin quantum effects suppress the Alfvén wave frequency in a hot and magnetized plasma.

Normally, for quantum effects to be significant in plasma, the temperature to density ratio is considered to be much small and the quantum effects are expected to be important on very small scales, i.e., debyelength $\lambda_D$ $= \left(kT/4\pi ne^2\right)^{1/2}$ and gyroradius $\rho_j$ should be smaller or of the order of the de Broglie wavelength $\lambda_{DB} = \hbar/(2\pi m/kT)^{1/2}$. At such small scale the quantum kinetic theory would be convenient for the proper treatment of these micro phenomenon. To study various aspects of quantum plasma, numerous efforts have been made whose applications range from plasmonic to dense astrophysical plasma [7]−[16].

Recently, there is much attention to study the dispersion effects of several low frequency modes by using the Bohm potential or quantum statistical effects and their practical implications which includes the Fermi pressure [17]. Various fluid plasma models have been developed to introduce the quantum effects through quantum corrections which either involve the quantum force produced by density fluctuations originating from the Bohm potential or through the spin of the particles which includes the magnetic dipole force and magnetization energy. At large scales, plasma is treated as an electrically conducting fluid and can be described by magnetohydrodynamics, while the situation has to be interpreted within the frame work of kinetic theory when microscopic scales are involved. The proper treatment of such phenomenon requires the development of kinetic theory based on quantum effects [18]. The presence of dust particles is known to significantly amend the electrostatic and electromagnetic modes [19]−[23] due to their presence through quasineutrality and responsible for some novel modes associated with dust particle dynamics and with the finite divergence of the cross-field plasma current density. More recently, Mahdavi and Azadboni investigated the non degenerate quantum effects on Weibel instability by considering its application in the absorption layer of fuel pallet, where thermal energy was taken larger than the Fermi energy with non negligible degeneracy and showed that instability is dependent on quantum effects [24].

In this paper, we have made an effort to examine the role of quantum diffraction effects associated with electrons in a dense, hot and magnetized plasma and its effect on low frequency modes like dust kinetic Alfvén wave instability. We report the existence and characteristics of short wavelength dust kinetic Alfvén waves by emphasizing on quantum diffraction effects arising through Bohm potential of electrons which is obviously dominant over the ions and the dust species due to large mass difference. The dispersion relation of quantum DKAW (QDKAW) is derived by incorporating Bohm quantum potential into the linearized Vlasov equation and the coupling of electrostatic quantum dust acoustic mode (QDAW) with kinetic Alfvén wave (KAW) and the theoretical aspect of KAW instability in an inertial regime is also discussed.

The manuscript is organized as follows: basic assumptions leading to the general dispersion relation are presented in sections 2 and 3 while results and conclusions are discussed in sections 4.

## 2 Basic assumptions

We consider an electromagnetic wave streaming instability in a collisionless electron-ion quantum dusty magnetoplasma. We assume strongly magnetized ions to be Maxwellian and a beam of non-degenerate electrons drifting across an external magnetic field ($\mathbf{B}_0 \| \hat{z}$) with constant ion drift velocity $V_0 \hat{x}$, i.e., ($V_0 \perp \mathbf{B}_0$), while dust is cold and unmagnetized. The plasma beta $\beta_i$ is assumed to be very small, i.e., $\beta_i \ll 1$, where $\beta_i = 4\pi n_{i0} T_i / B_0^2$. An electromagnetic wave with a wave vector $\mathbf{k}$ lies in $xz$ plane, $\mathbf{k} = k_\perp \hat{x} + k_\| \hat{z}$, making angle $\theta$ with $x-$plane, we may adopt two potential representation which is used in a low beta plasma to express the electromagnetic perturbations to describe the electric field $\mathbf{E}$, i.e., $E_\perp = -\nabla_\perp \varphi$, $E_\| = -\nabla_\| \psi$, where $\varphi \neq \psi$.

The linearized Poisson equation is

$$-(k_\perp^2 \varphi + k_\|^2 \psi) = 4\pi e[n_{e1} - n_{i1} - Z_{d0} n_{d1}], \tag{1}$$

where $n_{e1}$, $n_{i1}$, $n_{d1}$ are electron, ion and dust number densities respectively, $Q_{d0} = -Z_{d0} e$ (with $Z_{d0}$ as the number of electron charge on a grain) is the equilibrium charge on an average dust grain, and $e$ is the electron charge. For electromagnetic waves, we may ignore the dust charge fluctuation effects, i.e., $Z_{d1} = 0$ [25, 26].

Combining Ampere's and Faraday's law, we get, [27]

$$c^2 k_\perp^2 k_\| (\varphi - \psi) = 4\pi \omega \left[ J_{i1\|} + J_{d1\|} \right], \tag{2}$$

where $J's$ are the field aligned current densities for plasma species. Since the electrons are streaming perpendicular to the field direction, therefore, $J_{e1\|} = 0$.

The linearized Vlasov equation for quantum plasma by incorporating Bohm quantum potential in the direction of field can be written as

$$\frac{\partial f_{e1}}{\partial t} + v_\| \frac{\partial f_{e1}}{\partial z} + \left[ \frac{q_e}{m_e} E_z + \frac{\hbar^2}{4 m_e^2 n_0} \nabla \nabla^2 n_{e1} \right] \frac{\partial f_{e0}}{\partial v_\|} = 0.$$

The perturbed distribution function of streaming electrons is given by the aid of Vlasov equation and the zeroth order distribution function $f_{e0}$, which is the usual distribution function and is assumed to obey a Maxwellian or a Fermi Dirac distribution.

$$f_{e0} = A_e \exp\left(\frac{v_\|^2 + (v_\perp - V_0)^2}{v_{te}^2}\right), \tag{3}$$

where

$$A_e = n_{e0} \left(\frac{m_e}{2\pi T_e}\right)^{\frac{3}{2}}.$$

We again solve the Vlasov equation for hot and magnetized ions in terms of the guiding center coordinates and get for the perturbed distribution for any electromagnetic wave [28, 29].

$$f_{i1} = \left(\frac{n_{i0} e}{T_i}\right) \sum_l \sum_n \frac{k_\| v_\| \psi + n \Omega_{ci} \varphi}{\omega - n \Omega_{ci} - k_\| v_\|} \exp\left(i(n-l)\theta\right) J_n\left(\frac{k_\perp v_\perp}{\Omega_{ci}}\right) J_l\left(\frac{k_\perp v_\perp}{\Omega_{ci}}\right) f_{i0}, \tag{4}$$

where $f_{i0}$ is the equilibrium Maxwellian distribution function for ions and $\Omega_{ci}$ is the ion cyclotron frequency.

## 3 Basic theory and dispersion relation

By the aid of Vlasov equation the perturbed number density for electron including non-degenerate quantum effects can be recasted as

$$n_{e1} = \frac{2en_{e0}\varphi}{m_e v_{te}^2} \left( \frac{1 + \eta Z(\eta)}{1 + C_Q(1 + \eta Z(\eta))} \right), \quad (5)$$

where $C_Q = \hbar^2 k^3 / 4m_e^2 v_{te}^2 k_\parallel$, which shows the quantum correction in the number density of electrons and $Z(\eta)$ is plasma dispersion function for perpendicularly propagating electrons [30] with argument $\eta = (\omega - k_\perp V_0)/k_\perp v_{te}$.

The number density of hot and magnetized ions can be written as

$$n_{i1} = -\frac{n_{i0}e}{m_i} \frac{1}{k_\parallel v_{ti}^2} \sum_n \left[ k_\parallel v_{ti} \psi \left(1 + \xi_{in} Z(\xi_{in})\right) + n\Omega_{ci}\varphi Z(\xi_{in}) \right] I_n(b_i) e^{-b_i}, \quad (6)$$

where $I_n$ is the modified Bessel function with argument $b_i = k_\perp^2 v_{ti}^2 / 2\Omega_{ci}^2$ and $Z(\xi_{in})$ is the usual dispersion function for a Maxwellian plasma with $\xi_{in} = (\omega - n\Omega_{ci})/k_\parallel v_{ti}$.

The perturbed number density of cold and unmagnetized dust grains is given by using hyrodynamical model,

$$n_{d1} = \frac{n_{d0} Q_{d0}}{m_d \omega^2} \left( k_\perp^2 \varphi + k_\parallel^2 \psi \right). \quad (7)$$

Since the non-degenerate electrons are streaming along $x-$direction, therefore the longitudinal components of current density perturbations are taken to be zero, i.e., $J_{e1\parallel} = 0$. The parallel components current density for ions and dust species are

$$J_{i1\parallel} = -\frac{n_{i0}e^2}{T_i k_\parallel} \sum_n \left[ (1 + \xi_{in} Z(\xi_{in})) \left( k_\parallel v_{ti} \xi_{in} \psi + n\Omega_{ci}\varphi \right) \right] I_n(b_i) e^{-bi},$$

and

$$J_{d1\parallel} = \frac{n_{d0} Q_{d0}^2}{m_d \omega} k_\parallel \psi. \quad (8)$$

Now, using the explicit expressions of $n_{e1}, n_{i1}$ and $n_{d1}$ in the Eq. (1) and $J_{i1\parallel}, J_{e1\parallel}$ and $J_{d1\parallel}$ in the Eq. (2), allow to obtain the following system of equations,

$$A\varphi + B\psi = 0, \quad (9)$$
$$C\varphi + D\psi = 0,$$

where

$$A = k_\perp^2 + \frac{2\omega_{pe}^2}{v_{te}^2}\left(\frac{1+\eta Z(\eta)}{1+C_Q(1+\eta Z(\eta))}\right) + \frac{n\omega_{ci}}{k_\parallel}Z(\xi_{in})I_n(b_i)e^{-b_i} + k_\perp^2\frac{\omega_{pd}^2}{\omega^2},$$

$$B = k_\parallel^2 + \frac{2\omega_{pi}^2}{v_{ti}^2}\left(1+\xi_{io}Z(\xi_{io})\right)I_n e^{-b_i} - k_\parallel^2\frac{\omega_{pd}^2}{\omega^2},$$

$$C = c^2 k_\parallel k_\perp^2 + \frac{2\omega_{pi}^2}{v_{ti}^2}\frac{n\omega\Omega_{ci}}{k_\parallel}\left(1+\xi_{in}Z(\xi_{in})\right)I_n e^{-b_i}, \tag{10}$$

$$D = k_\parallel\left(k_\perp^2 c^2 + \omega_{pd}^2\right) - \omega\frac{1}{\lambda_{Di}^2}\sum_n \frac{v_{ti}}{2}\xi_{in}Z'(\xi_{in})I_n e^{-b_i}.$$

The dispersion relation is obtained by solving the homogeneous Eq. (9), i.e.,

$$AD - BC = 0. \tag{11}$$

Substituting the values of $A, B, C,$ and $D$ from Eq. (10), and after a straightforward algebra, we get

$$1 + \frac{2\omega_{pi}^2}{k_\parallel^2 v_{ti}^2}\sum_n\left[\left(1+\frac{\omega_{pd}^2}{k_\perp^2 c^2}\right)\frac{n\Omega_{ci}}{k_\parallel v_{ti}}Z(\xi_{in})I_n e^{-b_i} + \left(1+\xi_{in}Z(\xi_{in})\right)I_n e^{-b_i}\right] + \frac{2\omega_{pe}^2}{k_\parallel^2 v_{te}^2}\left[\left(1-\frac{\omega_{pd}^2}{k_\perp^2 c^2}\right)\left(\frac{1+\eta Z(\eta)}{1+C_Q(1+\eta Z(\eta))}\right)\right] + \left(1+\frac{\omega_{pd}^2}{k_\perp^2 c^2}\right)\frac{k_\perp^2}{k_\parallel^2}\left(1-\frac{\omega_{pd}^2}{\omega^2}\right) = 0, \tag{12}$$

which is the general dispersion relation of KAW streaming instabilities in a quantum dusty plasma. In the limit $\frac{\omega_{pd}^2}{k_\perp^2 c^2} \ll 1$, we get the dispersion relation of modified two stream instability in a dusty plasma [31] and in a dust less plasma the classical dispersion relation is obtained [32].

When the free energy associated with streaming is increased then $V_0 >> v_{th}$, where $v_{th}$ is the thermal speed of particles, the maximum growth rate also continues to increase and the dispersion equation for all plasma components often become nonresonant ( $\xi_{en,i} > 1$). The unstable mode in a non-resonant regime is typically known as fluid instability. In a cold plasma limit,

$$A = \frac{k_\perp^2 f_i}{\omega^2}\left(\omega^2 - \omega_{dlh}^2 - \frac{\omega^2 \omega_{pe}^2/f_i}{(\omega - k_\perp V_0)^2 - C_Q k_\perp^2 v_{te}^2}\right),$$

$$B = k_\perp^2 f_i - k_\parallel^2\frac{\omega_{pd}^2}{\omega^2},$$

$$C = c^2 k_\parallel k_\perp^2, \tag{13}$$

$$D = -k_\parallel\left(k_\perp^2 c^2 + \omega_{pd}^2\right).$$

where $\omega_{dlh}^2 = (\omega_{pd}^2 \Omega_{ci}^2)/\omega_{pi}^2$ and $f_i = \omega_{pi}^2/\Omega_{ci}^2$.

In order to investigate the behavior of instability in the presence of cross field streaming ions, we use of Eq. (11) and after few algebraic steps,

$$a\omega^4 + b\omega^3 + c\omega^2 + d\omega + e = 0, \tag{14}$$

$a = \alpha + k_\perp^2 c^2,$
$b = -2k_\perp^3 V_0 c^2 - 2k_\perp V_0 \alpha - 2k_\parallel^2 V_{Ai}^2 k_\perp V_0$
$c = \left(-\omega_{dlh}^2 - k_\parallel^2 V_{Ai}^2 C_Q - C_Q - k_\perp^2 V_0^2 - \omega_{pe}^2/f_i\right)\alpha - k_\parallel^2 V_{Ai}^2 \omega_{pd}^2 - C_Q k_\perp^2 c^2 - k_\perp^4 c^2 V_0^2$
$$\tag{15}$$
$d = 2k_\perp V_0 \omega_{dlh}^2 \alpha + 2k_\parallel^2 V_{Ai}^2 k_\perp V_0 \omega_{pd}^2$
$e = \left(C_Q - k_\perp^2 V_0^2\right)\omega_{dlh}^2 \alpha - k_\parallel^2 k_\perp^2 V_0 V_{Ai}^2 \omega_{pd}^2,$

where $\alpha = k_\perp^2 c^2 + \omega_{pd}^2$.

In the absence of streaming electrons the solution of Eq. (14) is given by

$$\omega^2 = \omega_{dlh}^2 + 2C_Q + \frac{k_\parallel^2 V_{Ai}^2}{\left(1 + k_\perp^2 \lambda_d^2\right)} \tag{16}$$

which is quantum corrected low frequency shear Alfvén wave in a dusty plasma, where $V_{Ai} = B_0/\sqrt{4\pi n_{i0} m_i}$ is the Alfvén speed and $\lambda_d = c/\omega_{pd}$, is the dust skin depth. In the limit $C_Q = 0$, we obtain the dispersion relation of KAW in inertial regime i.e., $\omega^2 = \omega_{dlh}^2 + k_\parallel^2 V_{Ai}^2/\left(1 + k_\perp^2 \lambda_d^2\right)$. When $C_Q = 0$, $k_\perp^2 \lambda_d^2 \ll 1$, we get the dust-modified dispersion relation of shear Alfvén waves, i.e., $\omega^2 = \omega_{dlh}^2 + k_\parallel^2 V_{Ai}^2$, which is a natural mode of any dusty magnetoplasma where $\omega_{dlh}^2$ provides constraint for the wave propagation. The analysis made here also indicates that cross-field streaming effects which are not coupled with perpendicular wavenumber, but appears as an additional term in the dispersion relation. The reason may be the absence of $J_{e\parallel}$ due to streaming in cross-field direction and the Bohm potential induced Vlasov model. It is therefore tempting to introduce quantum correction which may significantly modify the classical modes and related instabilities.

In order to observe the effect of dust grains on the wave in a low beta plasma, the limiting case of $k_\perp^2 \lambda_d^2 \gg 1$, can be obtained from the Eq.(16) in the form

$$\omega^2 = \omega_{dlh}^2 + \Omega_i \Omega_d \left(\frac{n_{d0} Z_{d0}}{n_{i0}}\right) \frac{k_\parallel^2}{k_\perp^2}. \tag{17}$$

The second term on R. H. S can be recognized as the convective cell frequency which strongly depends on dust parameters. It can be well recognized that dust inertia may play a major role in the wave dynamics, while the stationary dust excludes this mode. The heavy dust grains may be an important factor in diminishing the wave frequency and unstable regions of propagation.

## 4  Results

In the presented work, we have investigated the cross field streaming instability and dispersion properties of KAW due to non-degenerate electrons in a quantum dusty magnetoplasma. We have derived the KAW instability growth rates using various parameters close to dense astrophysical plasma, i.e., $n_{e0} = 10^{26} \text{cm}^{-3}, n_i = 1.001 \times 10^{26} \text{cm}^{-3}, Z_d = 10^3, n_{d0} = 0.3 \times 10^{23} \text{cm}^{-3}$, and $B_0 = 10^{13} G$. The influence of quantum Bohm potential is found to act against the KAW instability, while classical effects reinforce the unstable regions as depicted in Fig. 1. The cross field non-degenerate electron beam streaming with velocity $V_0$ helps to grow the unstable regions and a further increase inhibits the stabilization which maintains the wave amplitude as illustrated in Fig. 2. It has also been observed that quantum parameter $C_Q$ is inclined to suppress the instability. In Fig. 3, we have shown that the growth rates increase with the increase in number density, followed by increase in cross-filed ion streaming velocity. Physically, when the dust concentration is increased, it also enhances the depletion of electrons due to attachment to the dust grain surface, which inturns increases the streaming velocity which is ultimately responsible for the excitation of an electromagnetic wave. The role of dust charge is also found to enhance the wave activity which can be seen in Fig. 4. Our results indicate that KAWs in inertial regime are less affected by the Bohm potential effects and hence $C_Q$ appears to suppress the Alfvén wave frequency in a hot and magnetized non-degenerate quantum plasma which is evident from Fig. 5.

To summarize, we have investigated the effects of quantum contribution on the growth and dispersion of low frequency kinetic Alfvén wave in inertial regime by using Vlasov-Maxwell equations and have neglected the quantum effects of ions as they are heavier than electrons. It is found that quantum effects suppress the instability in a hot and magnetized plasma. The density fluctuations are not accompanied by pure electromagnetic non-streaming dense plasma, therefore, quantum correction make sense to electromagnetic waves when density fluctuations are involved especially in the inertial turbulent range. We therefore, have recasted the perturbed non-degenerate electron density which has a strong dependence on quantum correction parameter $C_Q$. Our analysis is based on linear approximation and we believe that we have contributed to the analysis of KAWI in a non-degenerate quantum plasma and there is a large set of nonlinear investigations which should be addressed. The present analysis could be applied to dense astrophysical streaming and ICF plasmas where quantum diffraction effects are non- negligible and significant.

- Figure 1. Contour plots between $\tilde{k}_\perp$ and $C_Q$ for $V_0 = 10$, $Z_d = 10^3$, $n_d = 10^{-8} n_i$
- Figure 2. Contour plots between $\tilde{k}_\perp$ and $\tilde{V}_0$ for $C_Q = 10$, $Z_d = 10^3$, $n_d = 10^{-8} n_i$
- Figure 3. Plots between $\tilde{k}$ and $\tilde{n}_d$ for $V_0 = 10$, $C_Q = 10$ and $Z_d = 10^3$.

- Figure 4. Effect of $Z_d$ on the dispersion characteristics.
- Figure 5. Effects of $C_Q$ on the real part of the dispersion relation.

# Conflict of interests

The authors declare that there is no conflict of interests regarding the publication of this paper.

# References

[1] Gary S Theory of Space Plasma Microinstabilities, Cambridge University Press, Cambridge, UK, (1993)

[2] Brinca A L, Romeiras F J, Gomberoff L J 2003 Geophys. Res. **108** (A1) 1038

[3] Chen L 2008 Plasma Phys. Control. Fusion **50** 124001

[4] Cramer F 2001 The Physics of Alfv´en Waves, WILEY-VCH Verlag Berlin GmbH, Berlin

[5] Gekelman W, Vincena S, Van Compernolle B, Morales G J, Maggs J, E Pribyl P and Carter T A 2011 Phys. Plasmas **18**, 055501

[6] Hussain A, Iqbal Z, Bordin G and Murtaza G 2013 Phys. Lett. A, **377**, 2131

[7] Haas F, Manfredi G, Feix M. R 2000 Phys. Rev. E 62, 2763

[8] Haas F, Garcia L G, Goedert J, Manfredi G 2003 Phys. Plasmas 10, 3858

[9] Haas F 2005 Phys. Plasmas 12, 062117

[10] Marklund M, Brodin G 2007 Phys. Rev. Lett. 98, 025001

[11] Marklund M, Eliasson B, Shukla P K 2007 Phys. Rev. E **76**, 067401

[12] Manfredi G 2005 Fields Inst. Commun. **46**, 263

[13] Brodin G, Marklund M 2007 Phys. Plasmas 14, 112107

[14] Brodin G, Marklund M, Eliasson B, Shukla P K 2007 Phys. Rev. Lett. **98**, 125001

[15] Shukla P K 2006 Phys. Lett. A **352**, 242

[16] Shukla P K and Stenflo L 2006 New J. Phys. **8**, 111

[17] Garcia L G, Haas F, de Oliviera L P L and Marklund J M 2005 Phys. Plasmas **12**, 082110

[18] Kumar S and Lu J Y 2012 Astrophys Space Sci, **341**, 597

[19] Rosenberg M and Shukla P K 2004 J. Plasma Physics **70,** 317

[20] Ali S and Shukla P K 2007 Eur. Phys. J. D **41,** 319

[21] Shukla P K and Mamun A A 2002 Introduction to Dusty Plasma Physics IOP, Bristol

[22] Choi C R and Lee D -Y 2007 Phys. Plasmas **14**, 052304

[23] Choi C R, Ryu C -M, Lee N C, and Lee D -Y 2005 Phys. Plasmas **12**, 022304

[24] Mahdavi M, Azadboni F K 2015 Advances in high energy physics, Article ID 746212

[25] Milandso F T, Aslaksen K and Havnes O 1993 Planet. Space. Sci. **42**, 321

[26] Salimullah M, Islam M K, Banerjee A K and Nambu M 2001 Phys. Plasmas **8**, 3510

[27] Hasegawa A and Chen L 1976 Phys. Fluids **19**, 1924

[28] Zubia K, Rubab N, Shah H A, Salimullah M and Murtaza G 2007 *Phys. Plasmas* **14,** 032105

[29] Liu C S and Tripathi V K 1986 *Phys. Rep.* **130**, 143

[30] Summers D and Thorne R M 1991 Phys. Fluids B **3**, 1835

[31] Rosenberg M and Krall N A 1995 Planet. Space Sci., **43**, 619

[32] Galeev A A and Sudan R N 1983 *Basic Plasma Physics*, edited by (Published by North-Holland Publishing Company, Amsterdam

Fig. 1

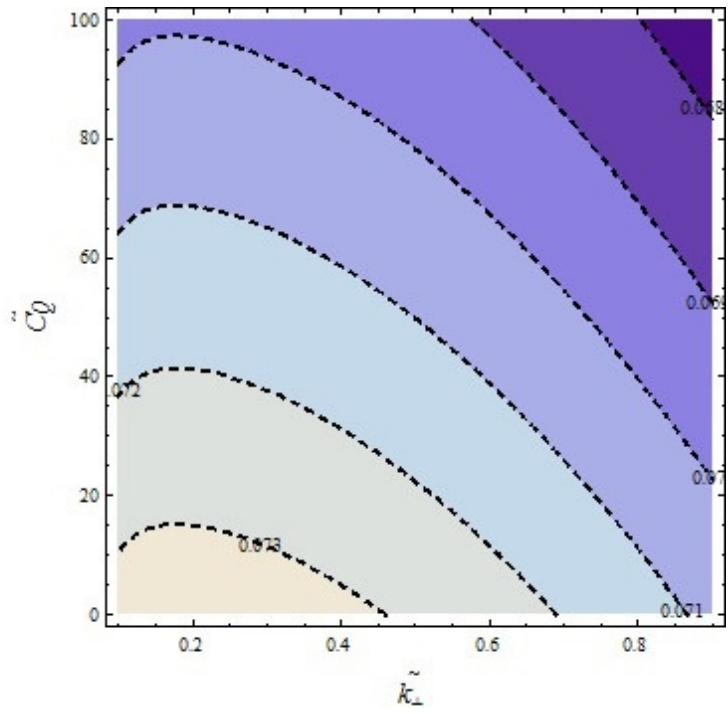

Fig. 2

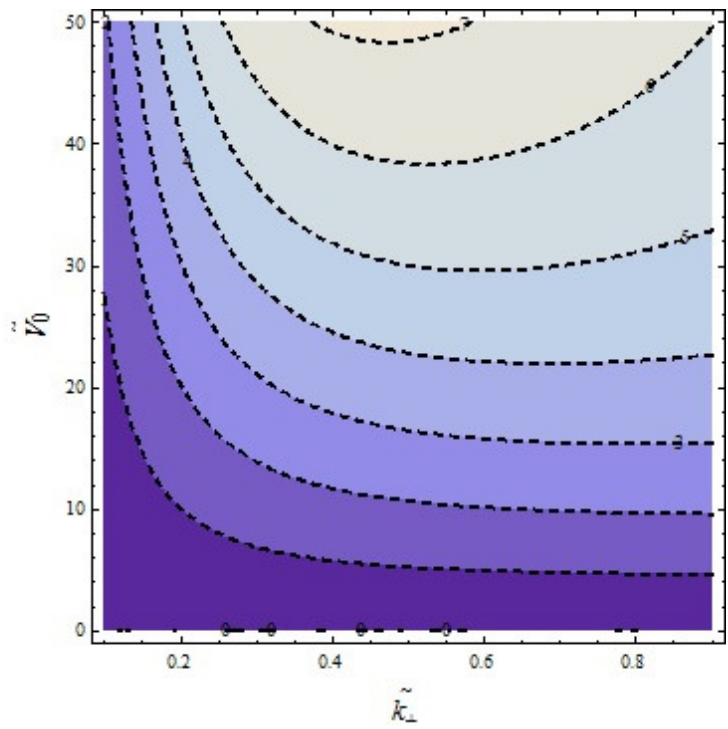

Fig. 3

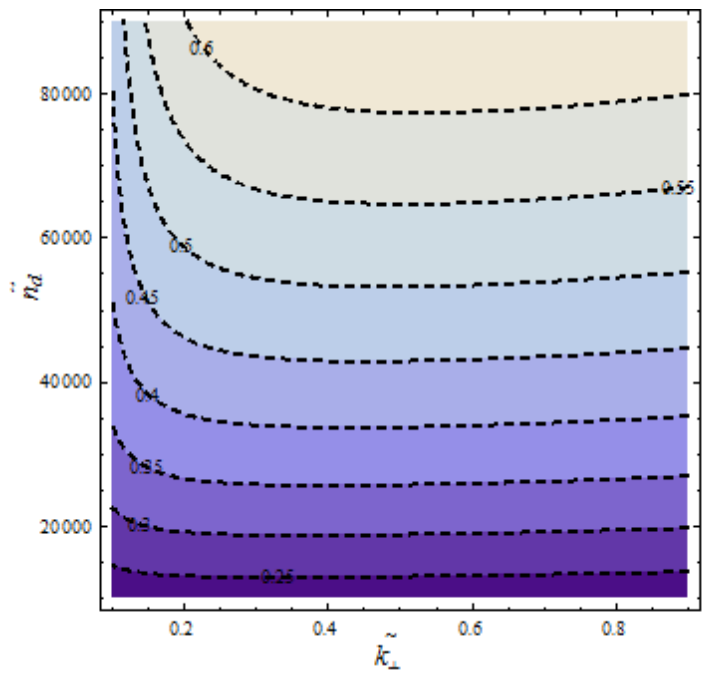

Fig. 4

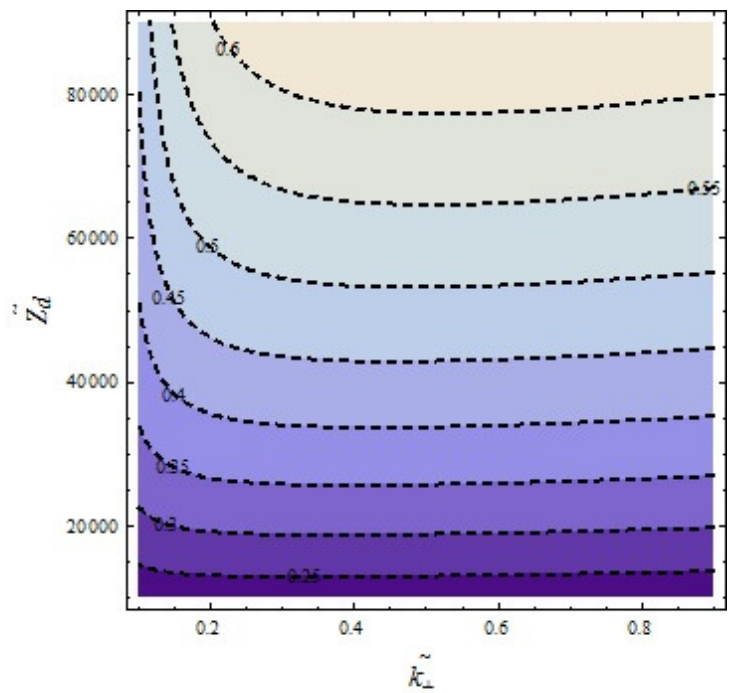

Fig. 5

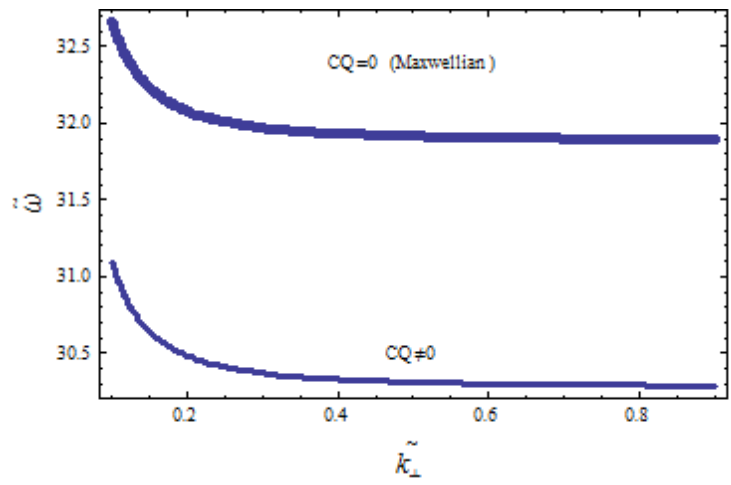